\documentclass[a4paper]{jpconf}
\usepackage{graphicx}
\usepackage{bm}
\begin{document}
\title{New ${\bm S}$=1/2 Kagom\'{e} Antiferromagnets A$_2$Cu$_3$SnF$_{12}$: \\
A=Cs and Rb}

\author{T. Ono$^1$, K. Morita$^1$, M. Yano$^1$, H. Tanaka$^1$, K. Fujii$^2$, H. Uekusa$^2$, Y. Narumi$^3$ and K. Kindo$^3$}

\address{$^1$Department of Physics, Tokyo Institute of Technology, Tokyo 152-8551, Japan}

\address{$^2$Department of Chemistry, Tokyo Institute of Technology, Tokyo 152-8551, Japan}

\address{$^3$Institute for Solid State Physics, The University of Tokyo, Chiba 277-8581, Japan}

\ead{o-toshio@lee.phys.titech.ac.jp}

\begin{abstract}
We synthesized single crystals of the new hexagonal compounds A$_2$Cu$_3$SnF$_{12}$ with A\,=\,Cs and Rb, and investigated their magnetic properties. These compounds are composed of Kagom\'{e} layers of corner-sharing CuF$_6$-octahedra. Cs$_2$Cu$_3$SnF$_{12}$ has the proper Kagom\'{e} layer at room temperature, and undergoes structural phase transition at $T_\mathrm{t}\,{\simeq}\,185$\,K. The temperature dependence of the magnetic susceptibility in Cs$_2$Cu$_3$SnF$_{12}$ agrees well with the result of the numerical calculation for $S\,{=}\,1/2$ two-dimensional Heisenberg Kagom\'{e} antiferromagnet down to $T_\mathrm{t}$ with the nearest exchange interaction $J/k_\mathrm{B}\,{\simeq}\,240$\,K. Although the magnetic susceptibility deviates from the calculated result below $T$\,$<$\,$T_\mathrm{t}$, the rounded maxima were observed at approximately $T\,{\simeq}\,(1/6)J/k_\mathrm{B}$ as predicted by the theory. Cs$_2$Cu$_3$SnF$_{12}$ undergoes three-dimensional magnetic ordering at $T_\mathrm{N}\,{=}\,20$\,K. 
Rb$_2$Cu$_3$SnF$_{12}$ has the Kagom\'{e} layer, whose unit cell is enlarged by $2a\,{\times}\,2a$ as compared with the proper Kagom\'{e} layer even at room temperature. From the viewpoint of crystal structure, the exchange interactions between nearest neighbor Cu$^{2+}$-ions are classified into four kinds. From the magnetic susceptibility and high-field magnetization measurements, it was found that the ground state is a disordered singlet with the spin gap, as predicted by recent theory. Exact diagonalization method with 12-site Kagom\'{e} cluster was performed to analyze the magnetic susceptibility. By comparing the calculated results with the experimental data, the individual exchange interactions were evaluated.
\end{abstract}

\section{Introduction}
Two-dimensional Heisenberg Kagom\'{e} antiferromagnet (2D HKAF) is of great interest from the viewpoint of the interplay of the frustration and quantum effects. 
For the case of $S$=1/2 2D HKAF, a disordered ground state was observed by various theoretical approaches\,\cite{Zeng,Sachdev,Chalker,Elstner,Nakamura}. Recent careful analyses and numerical calculations for an $S$\,=\,1/2 case demonstrated that the ground state is a spin liquid state composed of singlet dimers only, and that the ground state is gapped for triplet excitations, but gapless for singlet excitations\,\cite{Lecheminant,Waldtmann,Mambrini}. Consequently, magnetic susceptibility has a rounded maximum at $T\,{\sim}\,(1/6)J/k_{\rm B}$ and decreases exponentially toward zero with decreasing temperature, while specific heat exhibits a power law behavior at low temperatures \cite{Elstner,Misguich}. Specific heat also shows an additional structure, peak or shoulder at low temperatures after exhibiting a broad maximum at $T\,{\sim}\,(2/3)J/k_{\rm B}$. Here, we use the exchange constant $J$ defined as ${\cal H}_{\rm ex}\,{=}\,\sum_{\langle i,j\rangle} J_{ij}\,{\bm S}_i\cdot{\bm S}_j$\par
The experimental studies of the $S$=1/2 HKAF have been limited. The model substances include Cu$_3$V$_2$O$_7$(OH)$_2$$\cdot$2H$_2$O \cite{Hiroi}, $\beta$-Cu$_3$V$_2$O$_8$, \cite{Rogado} and [Cu$_3$(titmb)$_2$(CH$_3$CO$_2$)$_6$]$\cdot$H$_2$O\,\cite{Honda}. However, the above-mentioned intriguing predictions have not been verified experimentally. 
Recently, the herbertsmithite with the chemical formula ZnCu$_3$(OH)$_6$Cl$_2$ with the proper Kagom\'{e} lattice has been attracting considerable attention\,\cite{Shores,Mendels,Helton,Bert,Lee,Rigol,Misguich2}. Although no magnetic ordering occurs down to 50 mK\,\cite{Mendels}, magnetic susceptibility exhibits a rapid increase at low temperatures\,\cite{Helton,Bert}. This behavior was ascribed to a large number of idle spins ($4{\sim}10$\,\%) produced by intersite mixing between Cu$^{2+}$ and Zn$^{2+}$\,\cite{Bert,Lee,Misguich2,Olariu} and/or the Dzyaloshinsky-Moriya interaction\,\cite{Rigol}. The singlet ground state has not been observed experimentally. The search for new 2D HKAFs with $S$=1/2 is still ongoing.\par
Cs$_2$Cu$_3$ZrF$_{12}$ and Cs$_2$Cu$_3$HfF$_{12}$ have the proper Kagom\'{e} layer at room temperature\,\cite{Mueller} and are promising $S$=1/2 HKAFs\,\cite{Yamabe}. These systems undergo structural phase transitions at $T_{\rm t}\,{=}\,220$ and 170\,K, respectively, and also magnetic phase transitions at $T_{\rm N}\,{\simeq}\,24$ K\,\cite{Yamabe}. However, the magnetic susceptibilities observed at $T$\,$>$\,$T_{\rm t}$ can be perfectly described using theoretical results for an $S{=}1/2$ HKAF with rather large exchange interactions $J/k_{\rm B}\,{\sim}\,250$ K\,\cite{Morita}.

\begin{figure}[tbp]
	\begin{center}
		\includegraphics[width=14.2cm,clip]{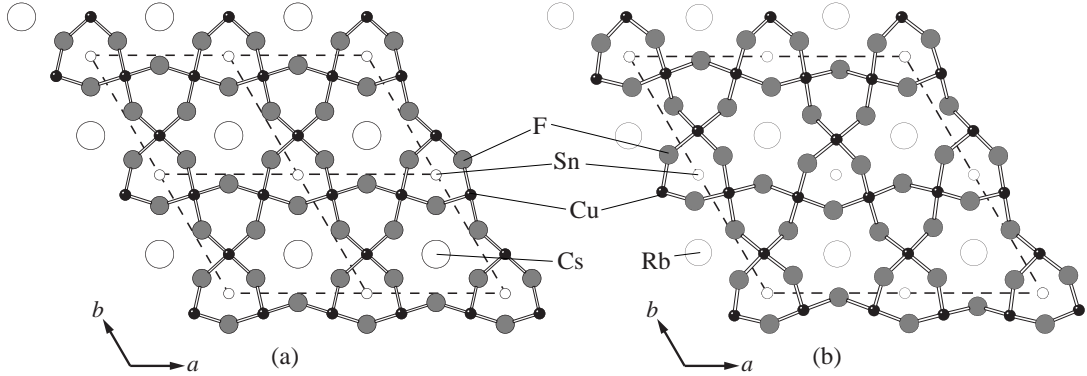}
	\end{center}
	\caption{Crystal structures of (a) Cs$_2$Cu$_3$SnF$_{12}$ and (b) Rb$_2$Cu$_3$SnF$_{12}$. Dashed lines denote the chemical unit cells in the $ab$ plane. Fluorine ions outside of the Kagom\'{e} layers are omitted to visualize the superexchange pathways between nearest Cu$^{2+}$-ions.}
	\label{Structure}
\end{figure}

In the present paper, we report the magnetic properties of the newly found Kagom\'{e} antiferromagnet Cs$_2$Cu$_3$SnF$_{12}$ together with Rb$_2$Cu$_3$SnF$_{12}$ which was already reported in the preceding paper\,\cite{Morita2008}. Figure\,\ref{Structure} shows the crystal structures of Cs$_2$Cu$_3$SnF$_{12}$ and Rb$_2$Cu$_3$SnF$_{12}$ viewed along the $c$-axes. 
In Fig {\ref{Structure}}, fluorine ions outside the Kagom\'{e} layers are omitted, so that the exchange pathways, Cu$^{2+}$\,$-$\,F$^{-}$\,$-$\,Cu$^{2+}$, are visible. The crystal structure of Cs$_2$Cu$_3$SnF$_{12}$ was found to be identical to those of Cs$_2$Cu$_3$$M$F$_{12}$ with $M$\,=\,Zr and Hf\,\cite{Mueller} with the lattice constants $a\,{=}\,7.142$\,\AA\ and $c\,{=}\,20.38$\,\AA\ at room temperature as shown in Fig.\,\ref{Structure} (a). The crystal structure of Rb$_2$Cu$_3$SnF$_{12}$ is closely related to that of Cs$_2$Cu$_3$SnF$_{12}$. The chemical unit cell is described by enlarging that of Cs$_2$Cu$_3$SnF$_{12}$ to $2a\,{\times}\,2a{\times}c$ as depicted in Fig.\,\ref{Structure} (b). Details of the crystal structure of Rb$_2$Cu$_3$SnF$_{12}$ is reported in ref.\,\cite{Morita2008}. In the present two systems, the magnetic Cu$^{2+}$-ions are surrounded by F$^-$-ions octahedrally and CuF$_6$ octahedra are linked in the $ab$-plane with sharing corners. Since CuF$_6$ octahedra for both systems are elongated along the principal axes that are approximately parallel to the $c$ axes, the hole orbitals $d(x^2\,{-}\,y^2)$ of Cu$^{2+}$-ions spread within the Kagom\'{e} layer. The sign and the magnitude of the exchange interactions between Cu$^{2+}$-ions are strongly dependant on the bond angles $\alpha$ of the Cu$^{2+}$\,$-$\,F$^{-}$\,$-$\,Cu$^{2+}$ pathways. The value of $\alpha$ for Cs$_2$Cu$_3$SnF$_{12}$ is ${\alpha}\,{=}\,139.9^{\circ}$ at room temperature. For Rb$_2$Cu$_3$SnF$_{12}$, there are four different values of $\alpha$ ($123.9^{\circ}\,{\sim}\,138.4^{\circ}$) producing four kinds of the exchange interactions. Details of the exchange interactions in Rb$_2$Cu$_3$SnF$_{12}$ are discussed in the last section. Since the angles $\alpha$ in the present systems are sufficiently far from 90$^{\circ}$ where the ferromagnetic superexchange occurs, the nearest-neighbor exchange interactions should be strongly antiferromagnetic. Furthermore, the Kagom\'{e} layers are well separated by non-magnetic Cs$^+$ or Rb$^+$, Sn$^{4+}$ and F$^+$ layers. Therefore, the present systems should be described as 2D $S$=1/2 HKAF.
\section{Experiments}
Single crystals of Cs$_2$Cu$_3$SnF$_{12}$ and Rb$_2$Cu$_3$SnF$_{12}$ were synthesized according to the chemical reaction $2\mathrm{AF} + 3\mathrm{CuF}_2 + \mathrm{SnF}_4$ $\rightarrow$ A$_2$Cu$_3$SnF$_{12}$ with A\,=\,Cs and Rb. The details of the sample preparation are described in ref.\,\cite{Morita2008}. The crystal structures at room temperature were analyzed using a Bruker SMART-1000 three-circle diffractometer equipped with a CCD area detector. Magnetic susceptibilities were measured in the temperature range 1.8$-$400 K using a SQUID magnetometer (Quantum Design MPMS XL). Magnetic fields were applied parallel and perpendicular to the $c$ axis. High-field magnetization measurement was performed for Rb$_2$Cu$_3$SnF$_{12}$ using an induction method with a multilayer pulse magnet at the Institute for Solid State Physics, The University of Tokyo.

\section{Results and Discussion}
\begin{figure}[tbp]
	\begin{center}
		\includegraphics[width=13.8cm,clip]{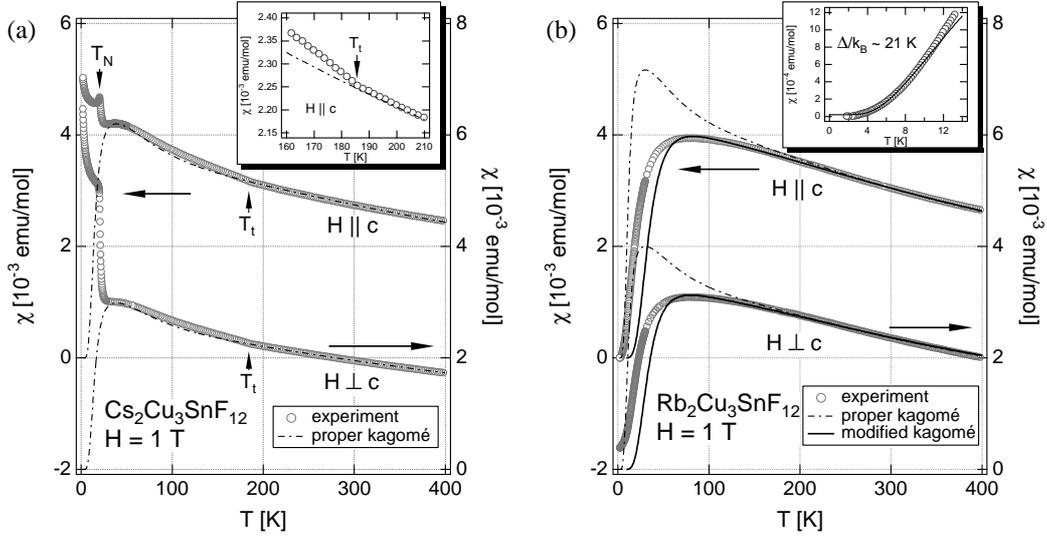}
	\end{center}
	\caption{Temperature variations of the magnetic susceptibilities in (a) Cs$_2$Cu$_3$SnF$_{12}$ and (b) Rb$_2$Cu$_3$SnF$_{12}$. Thin dot-dashed lines are the susceptibilities for the uniform KHAF obtained by the exact diagonalization for the 24-site Kagom\'{e} cluster\,\cite{Misguich2} with the parameters described in the text. Thick solid lines in (b) are the calculated susceptibilities for the modified Kagom\'{e} antiferromagnet described in the text. The inset of (a) shows the enlargement around the structural phase transition at $T_\mathrm{t}\,{\simeq}\,185$\,K. The inset of (b) denotes the low-temperature susceptibility for $H\,{\parallel}\,c$.}
	\label{Susceptibility}
\end{figure}

Figure\,\ref{Susceptibility} is the temperature variations of the magnetic susceptibilities $\chi$ of Cs$_2$Cu$_3$SnF$_{12}$ and Rb$_2$Cu$_3$SnF$_{12}$ for the field directions $H\,{\parallel}\,c$ and $H\,{\perp}\,c$ measured at $H\,{=}\,1$\,T. 
The susceptibilities of Cs$_2$Cu$_3$SnF$_{12}$ show small bend anomalies at $T_\mathrm{t}{\simeq}185$\,K due to the structural phase transition. At present, the details of crystal structure below $T$\,$<$\,$T_\mathrm{t}$ are not clear. The values of $\chi$ increases gradually with decreasing temperature, and shows the rounded maxima at $T_{\rm max}\,{\simeq}\,40$\,K for both field directions. At $T_\mathrm{N}\,{\simeq}\,20$\,K, $\chi$ for $H\,{\parallel}\,c$ shows $\lambda$-like anomaly indicative of the weak ferromagnetic moment along the $c$ axis. The sharp peak of $\chi$ at $T_\mathrm{N}$ is also observed in jarosite compound KFe$_3$(OH)$_6$(SO$_4$)$_2$\,\cite{Grohol}. In KFe$_3$(OH)$_6$(SO$_4$)$_2$, Kagom\'{e} layers which have the triangular spin structures with \textit{up} or \textit{down} weak ferromagnetic moment along the $c$ axis stack alternatively. The similar spin configurations are expected in Cs$_2$Cu$_3$SnF$_{12}$. The thin dot-dashed lines in Fig.\,\ref{Susceptibility} (a) are the theoretical susceptibilities for the $S{=}1/2$ uniform HKAF obtained by exact diagonalization method for 24-site Kagom\'{e} cluster\,\cite{Misguich2}\, with $J/k_\mathrm{B}\,{=}\,240$\,K and $g_\mathrm{{\parallel}c}\,{=}\,2.48$ for $H\,{\parallel}\,c$ and $g_\mathrm{{\perp}c}\,{=}\,2.09$ for $H\,{\perp}\,c$. For Cs$_2$Cu$_3$SnF$_{12}$, theoretical susceptibilities is in perfectly agreement with the experimental susceptibilities for $T$\,$>$\,$T_\mathrm{t}$. Although the experimental results slightly deviates from the theoretical result below $T$\,$<$\,$T_\mathrm{t}$, the temperature $T_{\rm max}$ at which the rounded maxima appear are approximately $(1/6)J/k_\mathrm{B}$ as predicted by the theory. \par
The experimental susceptibilities of Rb$_2$Cu$_3$SnF$_{12}$ in Fig.\,\ref{Susceptibility} (b) were corrected for small impurity effects according to the procedure described in ref.\,\cite{Morita2008}. With decreasing temperature, the susceptibilities exhibit rounded maxima at $T_{\rm max}\,{\sim}\,70$ K and shows the steep decrease. Any anomaly suggestive of the structural or magnetic phase transition was not observed between 1.8$-$400K. This result indicates clearly that the ground state is a disordered singlet with a spin gap, as predicted from a recent theory on a 2D $S$=1/2 HKAF. Assuming that Rb$_2$Cu$_3$SnF$_{12}$ can be described as the 2D spin gap system, the magnitude of the spin gap $\Delta$ was roughly estimated as $\Delta/k_\mathrm{B}\,{\simeq}\,21$\,K, fitting the relation $\chi\,{\propto}\,\exp(-\Delta/k_\mathrm{B}T)$\,\cite{Stone}\, to the data of $H\,{\parallel}\,c$ below $T$\,$<$\,$10$\,K as shown in the inset of Fig.\,\ref{Susceptibility} (b). 
\begin{figure}[tbp]
	\begin{minipage}{0.45\hsize}
		\begin{center}
			\includegraphics[width=50mm]{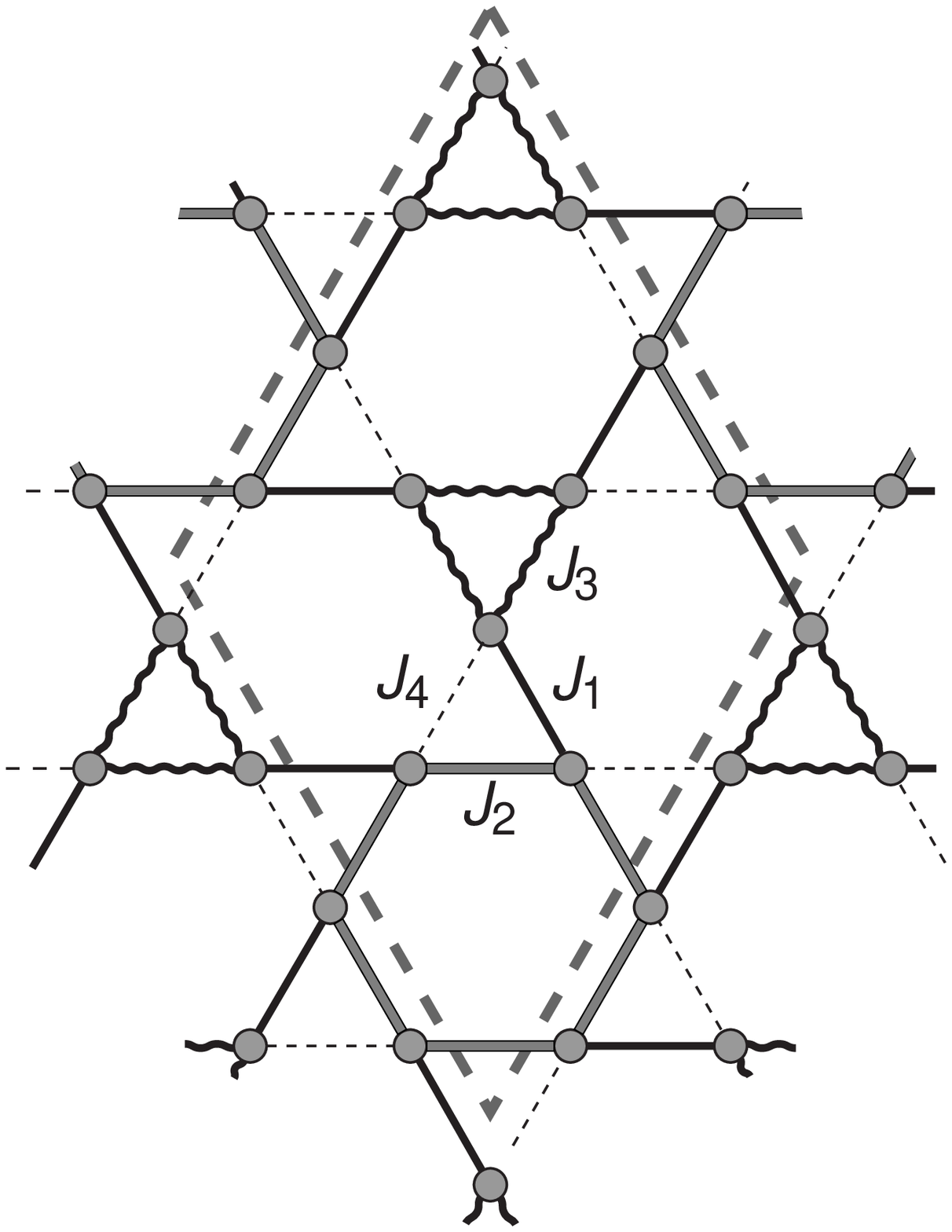}
		\end{center}
		\caption{Exchange network in the modified Kagom\'{e} layer of Rb$_2$Cu$_3$SnF$_{12}$. Thick dashed line denotes the chemical unit cell.}
		\label{exchange}
	\end{minipage}
	\hspace{0.05\hsize}%
	\begin{minipage}{0.5\hsize}
	\begin{center}
		\includegraphics[width=75mm]{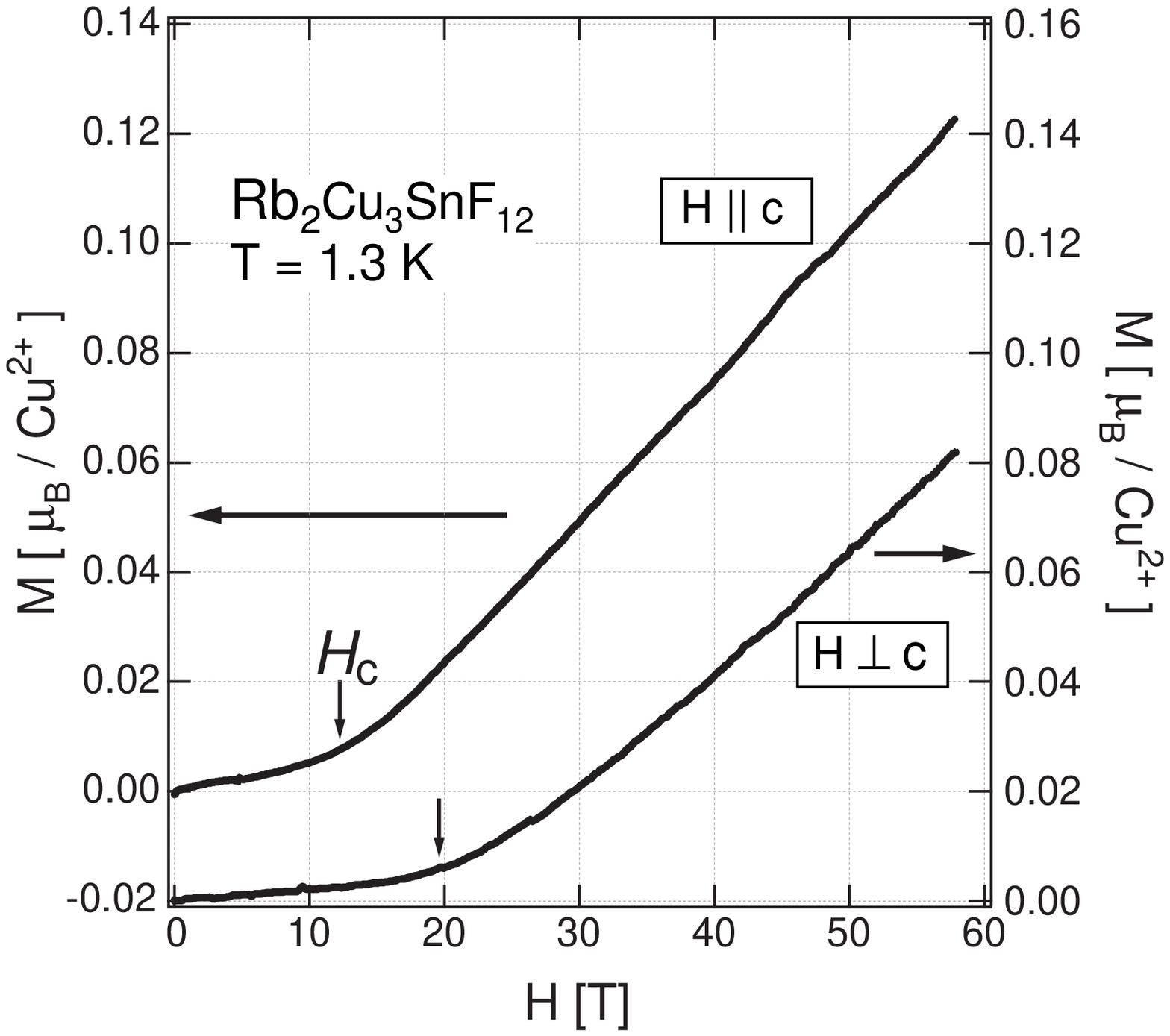}
	\end{center}
		\caption{Magnetization curves of Rb$_2$Cu$_3$SnF$_{12}$ measured at $T\,{=}\,1.3$ K for $H{\parallel}c$ and $H{\perp}c$. Arrows indicate the critical field $H_\mathrm{c}$.}
		\label{MH_curves}
	\end{minipage} 
\end{figure}
The susceptibility for $H\,{\parallel}\,c$ is almost zero for $T\,{\rightarrow}\,0$, whereas that for $H\,{\perp}\,c$ is finite. In Rb$_2$Cu$_3$SnF$_{12}$, since the elongated axes of CuF$_6$ octahedra incline alternately in the Kagom\'{e} layer, a staggered field should be induced when an external field is applied. Since there is no inversion center in the middle of two neighboring magnetic ions in the Kagom\'{e} layer, the Dzyaloshinsky-Moriya (DM) interaction is allowed. The Zeeman interaction due to the staggered field and the DM interaction can have finite matrix elements between the singlet ground state and the excited triplet state, because they are antisymmetric with respect to the interchange of the interacting spins. Thus, we infer that the ground state has a small amount of triplet component through these antisymmetric interactions when subjected to the external field parallel to the Kagom\'{e} layer. This gives rise to the finite susceptibility at $T$\,=\,0. 

For $T$\,$<$\,200 K, the susceptibility of Rb$_2$Cu$_3$SnF$_{12}$ does not agree with the theoretical result for the uniform $S{=}1/2$ KHAF with $J/k_\mathrm{B}{=}185$\,K and $g_\mathrm{\parallel c}\,{=}\,2.43$ and $g_\mathrm{\parallel c}\,{=}\,2.13$ as represented by the dot-dashed lines in Fig.\,\ref{Susceptibility} (b). This disagreement should be caused by the four different exchange interactions $J_1\,{\sim}\,J_4$ as depicted in Fig.\,\ref{exchange}. Since the chemical unit cell of Rb$_2$Cu$_3$SnF$_{12}$ has 12 spins, we carried out the exact diagonalization with 12-site Kagom\'{e} cluster. Since the antiferromagnetic exchange interaction becomes stronger with increasing bonding angle ${\alpha}$ of the exchange pathway Cu$^{2+}$\,$-$\,F$^{-}$\,$-$\,Cu$^{2+}$, the condition $J_1$\,$>$\,$J_2$\,$>$\,$J_3$\,$>$\,$J_4$ must be realized in Rb$_2$Cu$_3$SnF$_{12}$. Under this condition, we calculated susceptibility. As shown by thick solid lines in Fig.\,\ref{Susceptibility} (b), the best description can be obtained with the interactions $J_1\,{\sim}\,J_4$ listed in Table\,\ref{bonds} and the $g$-factors, $g_\mathrm{\parallel c}\,{=}\,2.43$ and $g_\mathrm{\parallel c}\,{=}\,2.13$. Calculated result describes well the experimental result for $T$\,$>$\,$T_\mathrm{max}$.
\begin{table}[tbp]
	\label{bonds}
	\caption{Bond angles $\alpha$ of the exchange pathway Cu$^{2+}$$-$F$^{-}$$-$Cu$^{2+}$ and corresponding exchange interactions in Rb$_2$Cu$_3$SnF$_{12}$ evaluated from the present analysis. The exchange parameters for Cs$_2$Cu$_3$SnF$_{12}$, Cs$2$Cu$3$ZrF$_{12}$\,\cite{Yamabe} and KCuGaF$_6$\,\cite{Morisaki} are also listed for comparison.}
	\begin{center}
		\begin{tabular}{ccccc}
			\br
				&Cu-Cu distance [\AA]& bond angle $\alpha$ [deg.]& $J/k_\mathrm{B}$ [K] &ref.\\
				\mr
				$J_1$	& 3.582 & 138.4 & 234 &   \\
				$J_2$	& 3.538 & 136.4 & 211 &   \\
				$J_3$	& 3.494 & 133.4 & 187 &   \\
				$J_4$	& 3.358 & 123.9 & 108 &   \\
				Cs$_2$Cu$_3$SnF$_{12}$& 3.571 & 139.9 & 240 &   \\
				Cs$_2$Cu$_3$ZrF$_{12}$& 3.583 & 141.6 & 244 & \cite{Yamabe}  \\
				KCuGaF$_6$& 3.395 & 129.1 & 103 & \cite{Morisaki} \\
				\br
			\end{tabular}
		\end{center}
\end{table}
For $T<T_\mathrm{max}$, the calculated susceptibility decreases more steeply than the experimental susceptibility. This should be ascribed to the finite-size effect.\par
We performed high-field magnetization measurements to evaluate the spin gap of Rb$_2$Cu$_3$SnF$_{12}$ directly. The results obtained $T$\,=\,1.3 K for $H\,{\parallel}\,c$ and $H\,{\perp}\,c$ are shown in Fig.\,\ref{MH_curves}. The magnetization is small up to the critical field $H_{\rm c}$ indicated by arrows and increases rapidly. The levels of the ground and excited states cross at $H_{\rm c}$. The magnetization anomaly at $H_{\rm c}$ is rather smeared. We infer that the antisymmetric interactions, such as the staggered Zeeman and DM interactions that can mix the triplet state into the singlet state, give rise to the smearing of the magnetization anomaly. We assign the critical field $H_\mathrm{c}$ to the field of inflection in $dM/dH$. The critical fields obtained for $H\,{\parallel}\,c$ and $H\,{\perp}\,c$ are $H_\mathrm{c}$\,=\,13(1)\,T and 20(1)\,T, respectively, which do not agree when normalized by the $g$-factor as $(g/2)H_\mathrm{c}$. When an external field is applied perpendicular to the $c$ axis, the magnetic susceptibility is finite even at $T$=0. Consequently, the ground state energy is not independent of the external field but decreases with the external field, resulting in an increase in the critical field. Therefore, as the spin gap, we take ${\Delta}/k_\mathrm{B}\,{=}\,21(1)$\,K obtained from $H_\mathrm{c}\,{=}\,13(1)$\,T for $H\,{\parallel}\,c$. This spin gap is the same as that evaluated from low-temperature susceptibility.

The present experiment revealed that Cs$_2$Cu$_3$SnF$_{12}$ undergoes 3D ordering, but Rb$_2$Cu$_3$SnF$_{12}$ does not. At present, we infer two candidates of origin leading to this difference. First candidate is the difference of the lattice distortion. Although crystal lattice of Cs$_2$Cu$_3$SnF$_{12}$ is distorted below $T_\mathrm{t}\,{=}\,185$\,K, the exchange network in Cs$_2$Cu$_3$SnF$_{12}$ is expected to be closer to the uniform case than that for Rb$_2$Cu$_3$SnF$_{12}$. When the exchange network becomes close to the uniform case, the spin gap should be smaller. Thus, we infer that the spin gap produced in a Kagom\'{e} layer of Cs$_2$Cu$_3$SnF$_{12}$ is so small as to be destroyed by the weak interlayer exchange, which leads to the 3D ordering. Second candidate is the DM interaction. C\'{e}pas \textit{et al}\,\cite{Cepas2008} claimed that even in the 2D-uniform $S=1/2$ Kagom\'{e} antiferromagnet, the long-range order can be triggered by the DM interaction. They argue that with increasing DM interaction, the system undergoes the quantum phase transition from the gapped state to the ordered state at $D_c \approx 0.1J$. The difference of the ground states between Cs$_2$Cu$_3$SnF$_{12}$ and Rb$_2$Cu$_3$SnF$_{12}$ may be produced by two origins mentioned above cooperatively, because the crystal structure and DM interaction is closely related with each other.

\section{Conclusion}
We have presented the results of magnetic measurements on the new hexagonal antiferromagnets Cs$_2$Cu$_3$SnF$_{12}$ and Rb$_2$Cu$_3$SnF$_{12}$. Magnetic susceptibilities in Cs$_2$Cu$_3$SnF$_{12}$ are well reproduced by the theoretical results for the uniform $S{=}1/2$ Heisenberg Kagom\'{e} antiferromagnet above the structural phase transition point $T$\,$>$\,$T_\mathrm{t}\,{=}\,185$\,K. The susceptibility shows a rounded maxima at approximately $T\,{\sim}\,(1/6)J/k_\mathrm{B}$ as predicted by the theory. However, Cs$_2$Cu$_3$SnF$_{12}$ undergoes 3D ordering at $T_\mathrm{N}\,{=}\,20$\,K. Rb$_2$Cu$_3$SnF$_{12}$ can be described as a modified Kagom\'{e} antiferromagnet with four kinds of neighboring exchange interaction. The results of magnetic susceptibility and high-field magnetization measurements revealed that the ground state is a disordeted singlet with a spin gap, as predicted from a recent theory. We have evaluated the individual exchange interactions using the exact diagonalization method with 12-site Kagom\'{e} cluster.

\ack
We express our sincere thank to G. Misguich for showing us his theoretical calculations. This work was supported by a Grant-in-Aid for Scientific Research from the Japan Society for the Promotion of Science, and by a 21 Century COE Program at Tokyo Tech ``Nanometer-Scale Quantum Physics'' and a Grant-in-Aid for Scientific Research on Priority Areas ``High Field Spin Science in 100 T'' both from the Japanese Ministry of Education, Culture, Sports, Science and Technology.

\end{document}